\documentclass[reprint,superscriptaddress,amsmath,amssymb,a4paperaps,pra,nofootinbib]{revtex4-1}
\usepackage{dcolumn,amssymb,amsmath,amsfonts,graphicx,latexsym,float,xcolor}
\usepackage{epstopdf}
\usepackage{nameref}
\usepackage{txfonts}
\usepackage{ulem}
\usepackage{easyReview}
\begin{document}

\title{Impurity induced quantum chaos for an ultracold bosonic ensemble in a double-well }
\author{Jie Chen}
\email {jie.chen@physnet.uni-hamburg.de}
\affiliation{Zentrum f\"ur Optische Quantentechnologien, Fachbereich Physik, Universit\"at Hamburg, Luruper Chaussee 149, 22761 Hamburg, Germany}
\author{Kevin Keiler}
\affiliation{Zentrum f\"ur Optische Quantentechnologien, Fachbereich Physik, Universit\"at Hamburg, Luruper Chaussee 149, 22761 Hamburg, Germany}
\author{Gao Xianlong}
\affiliation{Department of Physics, Zhejiang Normal University, Jinhua 321004, China}
\author{Peter Schmelcher}
\affiliation{Zentrum f\"ur Optische Quantentechnologien, Fachbereich Physik, Universit\"at Hamburg, Luruper Chaussee 149, 22761 Hamburg, Germany}
\affiliation{The Hamburg Centre for Ultrafast Imaging, Universit\"at Hamburg, Luruper Chaussee 149, 22761 Hamburg, Germany}
\date{\today}

\begin{abstract}
We demonstrate that an ultracold many-body bosonic ensemble confined in a one-dimensional (1D) double-well (DW) potential can exhibit chaotic dynamics due to the presence of a single impurity. The non-equilibrium dynamics is triggered by a quench of the impurity-Bose interaction and is illustrated via the evolution of the population imbalance for the bosons between the two wells. While the increase of the post-quench interaction strength always facilitates the irregular motion for the bosonic population imbalance, it becomes regular again when the impurity is initially populated in the highly excited states. Such an integrability to chaos (ITC) transition is fully captured by the transient dynamics of the corresponding linear entanglement entropy, whose infinite-time averaged value additionally characterizes the edge of the chaos and implies the existence of an effective Bose-Bose attraction induced by the impurity. In order to elucidate the physical origin for the observed ITC transition, we perform a detailed spectral analysis for the mixture with respect to both the energy spectrum as well as the eigenstates. Specifically, two distinguished spectral behaviors upon a variation of the interspecies interaction strength are observed. While the avoided level-crossings take place in the low-energy spectrum, the energy levels in the high-energy spectrum possess a band-like structure and are equidistant within each band. This leads to a significant delocalization of the low-lying eigenvectors which, in turn, accounts for the chaotic nature of the bosonic dynamics. By contrast, those highly excited states bear a high resemblance to the non-interacting integrable basis, which explains for the recovery of the integrability for the bosonic species. Finally, we discuss the induced Bose-Bose attraction as well as its impact on the bosonic dynamics.
 
\end{abstract}

\maketitle
\section{Introduction}
Trapping of an ultracold many-body bosonic ensemble in a one-dimensional (1D) double-well (DW) potential constitutes a prototype system for the investigations of the correlated quantum dynamics  \cite{DW_exp_1,DW_exp_2, DW_exp_3}. Such a system represents a bosonic Josephson junction (BJJ), an atomic analogy of the Josephson effect initially predicted for Cooper pair tunneling through two weakly linked superconductors  \cite{BJJ_1,BJJ_2}. Owing to the unprecedented controllability of the trapping geometries as well as the atomic interaction strengths \cite{cold_atom_rev}, studies of the BJJ unveil various intriguing phenomena which are not accessible for conventional superconducting systems \cite{BJJ_Rabi_1, BJJ_Rabi_2, BJJ_Rabi_3, BJJ_Frag_1,BJJ_Frag_2, BJJ_Squeeze_1, BJJ_Squeeze_2}. Examples are the Josephson oscillations \cite{BJJ_Rabi_1, BJJ_Rabi_2, BJJ_Rabi_3}, fragmentations  \cite{BJJ_Frag_1, BJJ_Frag_2}, macroscopic quantum self trapping \cite{DW_exp_3, BJJ_Rabi_1, BJJ_Rabi_2},  collapse and revival sequences \cite{BJJ_Rabi_3} as well as the atomic squeezing state \cite{BJJ_Squeeze_1, BJJ_Squeeze_2}. 

Under the explicit time-dependent driving forces, the BJJ can alternatively turn into the quantum kicked top (QKT), a famous platform for the investigations of quantum chaos as well as the classical-quantum correspondence \cite{QKT_1, QKT_2, QKT_3, QKT_4, QKT_5, QKT_6, QKT_7, QKT_8, QKT_9, QKT_10, QKT_11, QKT_12, QKT_13}. To date, related studies include the spectral statistics \cite{QKT_2}, the entanglement entropy production \cite{QKT_3, QKT_4, QKT_5, QKT_6, QKT_7, QKT_8, QKT_9, QKT_10}, the quantum decoherence and quantum correlations \cite{QKT_11, QKT_12} as well as the border between regular and chaotic dynamics \cite{QKT_13}. Moreover, by viewing the QKT as a collective $N$-qubit system, the effects of the quantum chaos on the digital quantum simulations have also been detailed discussed recently \cite{QKT_14,QKT_15}. 

On the other hand, stimulated by the experimental progresses on few-body ensembles \cite{few_exp1,few_exp2, few_exp3, few_exp4, few_exp5,few_exp6}, significant theoretical effort also focuses on the 1D few-body atomic systems \cite{few_gs_1, few_gs_2, few_gs_3, few_gs_4, few_gs_5, few_gs_6, few_gs_7, few_quench_1, few_quench_2, few_quench_3, few_bf_SC1, few_bf_SC2}, revealing for example the ground state \cite{few_gs_1, few_gs_2, few_gs_3, few_gs_4, few_gs_5, few_gs_6, few_gs_7, few_bf_SC1, few_bf_SC2} as well as the dynamical properties \cite{few_quench_1, few_quench_2, few_quench_3}, which pave the way for the studies of the binary mixtures with large particle number imbalance. Such hybridized systems are deeply related to the polaron physics \cite{polaron_1, polaron_2, polaron_3} as well as the open quantum systems \cite{OQS_1} and are particularly interesting owing to the fact that one subsystem is in the deep quantum regime while the other one can more or less be described by the semi-classical physics. Note, however, that while most of the discussions focus on impacts on the minority species from the majority bath, studies which alternatively explore the feedback to the majority species due to the presence of the minority one are still rare.

In the present paper, we investigate a binary ultracold atomic mixture made of a single impurity and a non-interacting many-body bosonic ensemble that are confined within a 1D DW potential. Unlike most of the previous studies where the focuses are put on the weak-interacting regime, rendering the impurity being restricted into the lowest two modes of the DW potential \cite{Impurity_BH_1, Impurity_BH_2, Impurity_BH_3, Impurity_BH_4, Impurity_BH_5, Impurity_BH_6}, our discussions are not restricted to such a scenario. Specifically, we study the onset of the chaos for the majority bosonic species due to the presence of the impurity and put particular emphasis on the its dynamical response upon a sudden quench of the impurity-Bose interaction strength. As an exemplary observable, we monitor the quantum evolution of the population imbalance for the bosons between the two wells starting from a balanced particle population. While the increase of the post-quench interaction strength always facilitates a chaotic motion for the bosonic population imbalance, it becomes regular again when the impurity initially is prepared in the highly excited states. In order to characterize such an integrability to chaos (ITC) transition, we employ the linear entanglement entropy as a signature of quantum chaos, which alternatively measures the decoherence for the bosonic species. Depending on the degree of chaos, the transient dynamics of the corresponding linear entanglement entropy can behave as either a rapid growth or a slow variation with increasing time, whereas, its infinite-time averaged value, in addition, captures the edge of quantum chaos, i.e., the border between the integrable and the chaotic regions in the corresponding classical phase space. Furthermore, by computing the infinite-time averaged values of the linear entanglement entropy for various initial conditions, we find a striking resemblance between its profile and a classical phase space with attractive Bose-Bose interaction, which implies the existence of an attractive interaction among the bosons induced by the impurity. 

In order to elucidate the physical origin for the above observed ITC transition, we perform a detailed spectral analysis with respect to both the energy spectrum as well as the eigenstates of the mixture. Two distinguished spectral behaviors upon a variation of the interspecies interaction strength are observed. While the avoided level-crossings take place in the low-energy spectrum, the energy levels in the high-energy spectrum possess a band-like structure and are equidistant within each band. Consequently, this results in a significant delocalization for those low-lying eigenstates which, in turn, accounts for the chaotic nature of the bosonic non-equilibrium dynamics. Remarkably, those highly excited states bear a striking resemblance to the non-interacting integrable basis, which explains the recovery of the integrability for the bosonic species. Finally, we also discuss the induced Bose-Bose attraction and its impact on the bosonic dynamics.

This paper is organized as follows. In Sec.~\ref{Setup}, we introduce our setup including the Hamiltonian, the initial conditions as well as the quantities of interests. In Sec.~\ref{Results}, we present our main observation: the ITC transition for the bosonic species. In Sec.~\ref{Discussions}, we perform a detailed spectral analysis for the mixture with respect to both the energy spectrum as well as the eigenstates, so as to elucidate the physical origin for the above observed ITC transition. Finally, our conclusions and outlook are provided in Sec.~\ref{Conclusions}.  

\section{Setup} \label{Setup}
\subsection{Hamiltonian and angular-momentum representation} \label{Hamiltonian}
The Hamiltonian of our 1D ultracold impurity-Bose mixture is given by $\hat{H} = \hat{H}_{I} + \hat{H}_{B} +  \hat{H}_{IB}$, where 
\begin{align}
\hat{H}_{\sigma} &=\int dx_{\sigma}~\hat{\psi}^{\dagger}_{\sigma}(x_{\sigma}) \textit{h}_{\sigma}(x_{\sigma}) \hat{\psi}_{\sigma}(x_{\sigma}), \nonumber\\
\hat{H}_{IB} &= {g_{IB}}\int dx~\hat{\psi}^{\dagger}_{I}(x) \hat{\psi}^{\dagger}_{B}(x) \hat{\psi}_{B}(x) \hat{\psi}_{I}(x), \label{Hamiltonian_IB}
\end{align}
and $\textit{h}_{\sigma}(x_{\sigma}) = -\frac{\hbar^{2}}{2 m_{\sigma}}\frac{\partial^{2}}{\partial x_{\sigma}^{2}}+ V_{DW}(x_{\sigma})$ is the single-particle Hamiltonian for the $\sigma = I(B)$ species being confined within a 1D symmetric DW potential $V_{DW}(x_{\sigma}) = a_{\sigma} (x_{\sigma}^{2} - b_{\sigma}^{2})^{2}$. For simplicity, we consider the atoms for both species are of the same mass ($m_I = m_B = m$) and are trapped by the same potential geometry, i.e., $a_I = a_B = a_{DW}$ and $b_I = b_B = b_{DW}$.
$ \hat{\psi}_{\sigma}^{\dagger}(x_{\sigma})$ [$\hat{\psi}_{\sigma}(x_{\sigma})$] is the field operator that creates (annihilates) a $\sigma$-species particle at position $x_{\sigma}$. Moreover, we neglect the interactions among the bosons and assume the impurity-Bose interaction is of zero range and can be modeled by a contact potential of strength \cite{Feshbach_1, Feshbach_2, Feshbach_3,few_quench_3}
\begin{equation}
g_{IB} = \frac{2 \hbar^{2}a_{3D}}{\mu a_{\bot}^{2}}[1-C \frac{a_{3D}}{a_{\bot}}]^{-1}.
\end{equation}
Here $a_{3D}$ is the 3D impurity-Bose $s$-wave scattering length and $C \approx 1.4603$ is a constant. The parameter $a_{\bot} = \sqrt{ \hbar/ \mu \omega_{\bot}}$ describes the transverse confinement with $\mu = m/2$ being the reduced mass and we assume the transverse trapping frequency $\omega_{\bot}$ to be equal for both species. In the following discussions, we rescale the Hamiltonian of the mixture $\hat{H}$ for the units of the energy, length and time as $\eta = \hbar  \omega_{\bot}$, $\xi = \sqrt{\hbar /m  \omega_{\bot}}$ and $\tau = 1/\omega_{\bot}$, respectively. We focus on the repulsive interaction regime, i.e., $g_{IB} \geqslant 0$ and set $a_{DW} = 0.5$, $b_{DW} = 1.5$, such that the lowest two single-particle energy levels are well separated from the others [see Fig.~\ref{ps} (a), the spatial geometry of $V_{DW}(x)$ (black dashed line) as well as the lowest six single-particle energy levels (grey solid lines)]. Throughout this work, we explore a binary mixture made of a single impurity and 100 bosons ($N_{I}=1$, $N_{B} = 100$), and focus on the dynamical response for the majority bosonic species upon a sudden quench of the impurity-Bose interaction strength (see below). Let us note that such a 1D mixture is experimentally accessible by imposing strong transverse and weak longitudinal confinement for a binary e.g., Bose-Fermi mixture with two different kinds of atoms \cite{mixture_exp_bf_1, mixture_exp_bf_2} or a Bose-Bose mixture made of the same atoms with two different hyperfine states \cite{mixture_exp_bb_1, mixture_exp_bb_2}. The DW potential can also be readily constructed by imposing a 1D optical lattice on top of a harmonic trap \cite{DW_exp_3, BJJ_2}. Moreover, the contact interaction strength $g_{IB}$ can be controlled experimentally by tuning the $s$-wave scattering lengths via Feshbach or confinement-induced resonances \cite{Feshbach_1,Feshbach_2,Feshbach_3}.

Noticing further that the bosonic species is confined within a tight DW potential with $\delta_{1} \gg \delta_{0}$ [c.f. Fig.~\ref{ps} (a)], here $\delta_{i}$ denotes the energy difference between the $i$-th and the $(i+1)$-th single-particle eigenstates. We adopt the two-mode approximation
\begin{equation}
\hat{\psi}_{B}(x) = u_{L}(x) \hat{b}_{L} + u_{R}(x) \hat{b}_{R}, \label{2_mode_psi}
\end{equation}
with $u_{L,R}(x)$ being the Wannier-like states localized in the left and right well, respectively. This leads to the low-energy effective Hamiltonian for the bosonic species
\begin{equation}
\hat{H}_{B} =  -J_{0} (\hat{b}^{\dagger}_{L}\hat{b}_{R} + \hat{b}^{\dagger}_{R} \hat{b}_{L} ), \label{BH_model}
\end{equation}
corresponding to the two-site Bose-Hubbard (BH) model with $J_{0} = 0.071$ being the hopping amplitude.

Before proceeding, it is instructive to express the above BH Hamiltonian in the angular-momentum representation. To see this, we introduce three angular-momentum operators \cite{BJJ_Rabi_3, BJJ_4}
\begin{align}
\hat{J}_{x} &= \frac{1}{2} (\hat{b}_{L}^{\dagger} \hat{b}_{R}  + \hat{b}_{R}^{\dagger} \hat{b}_{L} ) ,~~~\hat{J}_{y} = -\frac{i}{2} (\hat{b}_{L}^{\dagger} \hat{b}_{R}  - \hat{b}_{R}^{\dagger} \hat{b}_{L} ), \nonumber\\
\hat{J}_{z} &= \frac{1}{2} (\hat{b}_{L}^{\dagger} \hat{b}_{L}  - \hat{b}_{R}^{\dagger} \hat{b}_{R} ),  \label{spin_operators}
\end{align}
obeying the SU(2) commutation relation $ [\hat{J}_{\alpha}, \hat{J}_{\beta}] = i \epsilon_{\alpha \beta \gamma} \hat{J}_{\gamma}$. The BH Hamiltonian in Eq.~\eqref{BH_model} thus can be rewritten as 
\begin{equation}
\hat{H}_{B} = - 2J_0 \hat{J}_{x}, \label{spin_BH_model}
\end{equation}
which describes the angular momentum precession of a single particle whose spatial degrees of freedom (DOFs) are frozen. According to definitions for $\hat{J}_{x}$ and $\hat{J}_{z}$ in Eq.~\eqref{spin_operators}, we note that the kinetic energy in the BH model as well as the population imbalance for the bosons between the two wells are in analogy to the magnetizations of this single particle along the $x$ and the $z$ axes. Moreover, the particle number conservation in the Hamiltonian \eqref{BH_model} corresponds to the angular momentum conservation 
\begin{equation}
\hat{J}^{2} = \hat{J}_{x}^{2} + \hat{J}_{y}^{2} + \hat{J}_{z}^{2} = \frac{N_{B}}{2} (\frac{N_{B}}{2}+1)
\end{equation}  
for the Hamiltonian \eqref{spin_BH_model}.  

For the case $g_{IB} = 0$, the angular momentum dynamics can be simply integrated out from the corresponding Heisenberg equations of motion, in which 
\begin{align}
\hat{J}_{y}(t) &= \hat{J}_{z}(0) \text{cos}(2J_{0}t) - \hat{J}_{y}(0) \text{sin}(2J_{0}t),  \nonumber \\
\hat{J}_{z}(t) &= \hat{J}_{y}(0) \text{cos}(2J_{0}t) + \hat{J}_{z}(0) \text{sin}(2J_{0}t), \label{EOM_Jyz}
\end{align}
being the harmonic oscillations with the frequency $\omega_0 = 2J_{0}$ and $\hat{J}_{x} (t) = \hat{J}_{x} (0)$ is time-independent since $[\hat{J}_{x}, \hat{H}_{B}] = 0$. Further introducing the normalized vector $\hat{\vec{S}}(t) = \hat{S}_{x}(t) \vec{i} + \hat{S}_{y}(t) \vec{j}+ \hat{S}_{z}(t) \vec{k}$ with $ \hat{S}_{\gamma}(t) = \hat{J}_{\gamma}(t)/J$ for $\gamma = x, y, z$ and $J= N_{B}/2$, together with the fact that
\begin{equation}
\sum_{\gamma = x,y,z}\langle\hat{S}_{\gamma}\rangle^{2}(t)   = \sum_{\gamma = x,y,z} \langle\hat{S}_{\gamma}\rangle^{2}(0), \label{normal_S}
\end{equation}
one can readily show that the motion of the vector $\hat{\vec{S}}(t)$ always lies on the Bloch sphere with unit radius if, in addition, we choose the initial state as the atomic coherent state (ACS) (see below).

\subsection{Classical dynamics} \label{Classical_dynamics}
The above angular momentum dynamics can alternatively be understood in a classical manner. As we will show below, the periodic motions for $\hat{J}_{y} (t)$ and $\hat{J}_{z} (t)$ [equivalently $\hat{S}_{y} (t)$ and $\hat{S}_{z} (t)$] correspond to the periodic oscillation of a classical non-rigid pendulum around its equilibrium position, while the conservation of $\hat{J}_{x} (t)$ [$\hat{S}_{x} (t)$] relates to the energy conservation of this pendulum. To this end, we first adopt the mean-field approximation as $\hat{b}_{\beta} = b_{\beta}$ ($ \beta = L,R$) with $b_{\beta} $ being a $c$-number \cite{GPE_1}. The quantum operators $\hat{S}_{x}$, $\hat{S}_{y}$ and $\hat{S}_{z}$ then should be rewritten as 
\begin{align}
S_{x} &= \frac{1}{2J} (b_{L}^{\ast} b_{R}  + b_{R}^{\ast} b_{L} ), ~~~ S_{y} = -\frac{i}{2J} (b_{L}^{\ast} b_{R}  - b_{R}^{\ast} b_{L} ), \nonumber \\
S_{z} &= \frac{1}{2J} (b_{L}^{\ast} b_{L}  - b_{R}^{\ast} b_{R} ).  \label{spin_classical}
\end{align}
Employing the phase-density representation for $b_{\beta}$ as $b_{\beta} = \sqrt{N_{\beta}^{B}} e^{i \theta_{\beta}}$ and further introducing the two conjugate variables 
\begin{equation}
Z = (N_{L}^{B} - N_{R}^{B}) /N_B,~~~~~~\varphi = \theta_{R} - \theta_{L} \label{cl_z_phi},
\end{equation}
representing the relative population imbalance between the two wells and the relative phase difference, respectively, we arrive at 
\begin{equation}
S_{x} = \sqrt{1-Z^{2}} \text{cos}\varphi, ~~~ S_{y} = \sqrt{1-Z^{2}} \text{sin}\varphi, ~~~S_{z} = Z,  \label{spin_classical_Z_phi}
\end{equation} 
whose dynamics are governed by the Hamiltonian 
\begin{equation}
H_{cl} = - J_{0}\sqrt{1-Z^{2}} \text{cos}\varphi, \label{classical_pendulum}
\end{equation}
which, as aforementioned, describes a non-rigid pendulum with angular momentum $Z$ whose length is proportional to $\sqrt{1-Z^{2}} $ \cite{BJJ_Rabi_1, BJJ_Rabi_2, BJJ_Rabi_3, BJJ_Driven_1}. Comparing the Eq.~\eqref{spin_classical_Z_phi} to the Eq.~\eqref{classical_pendulum}, we note that $S_{y}$ and $S_{z}$, being the classical counterpart of the quantum operators $\hat{S}_{y}$ and $\hat{S}_{z}$, now represent the horizontal displacement and the angular momentum of this classical pendulum, while the $S_{x}$ ($\hat{S}_{x}$) proportions to its total energy which is conserved during the dynamics. 

\begin{figure}
  \centering
  \includegraphics[width=0.5\textwidth]{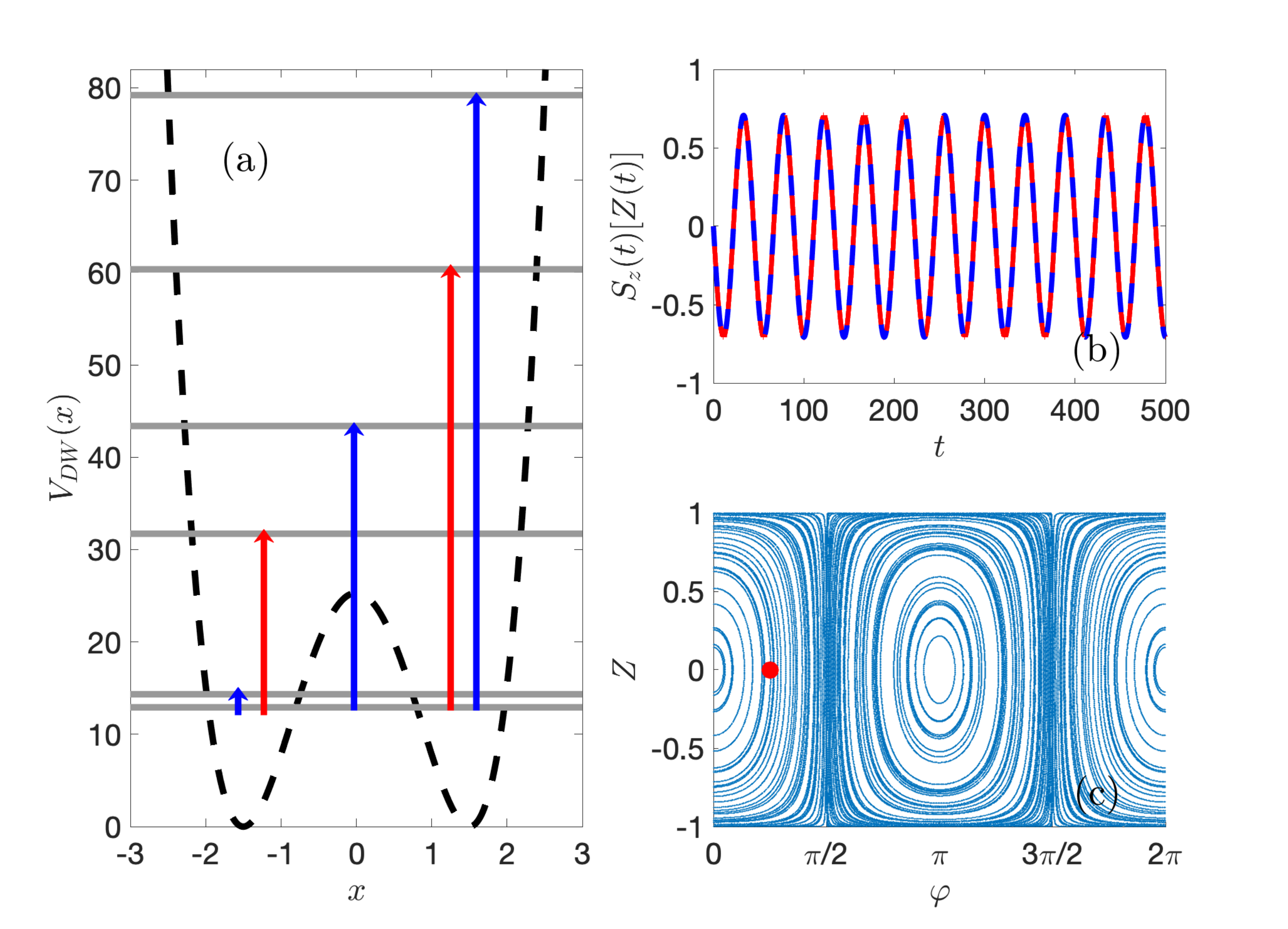}\hfill
  \caption{(Color online) (a) Single-particle spectrum for the double-well potential, in which the gray horizontal lines denote the lowest six energy levels and the blue (red) arrows represent possible transitions that reverse (preserve) the spatial parity of the impurity. (b) Real-time dynamics for the bosonic population imbalance $S_{z}(t)$ for the initial state $|\Psi (0)\rangle = |\phi_{0} \rangle \otimes |\pi/2, \pi/4 \rangle$ and for the case $g_{IB} = 0$ (red solid line), together with the classical $Z(t)$ dynamics starting from the phase point $(Z = 0, \varphi = \pi/4 )$ (blue dashed line). (c) Classical phase space for $J_{0} = 0.071$. The red dot denotes the phase space point ($Z = 0, \varphi = \pi/4$) corresponding to the ACS $|\theta, \varphi \rangle = |\pi/2, \pi/4 \rangle$.}
\label{ps}
\end{figure}

In this way, an one-to-one correspondence between the quantum and classical dynamics is established in which the periodic motions for $\hat{S}_{y} (t)$ and $\hat{S}_{z} (t)$ are mapped to the periodic oscillations for this classical pendulum around its equilibrium position. Since our focus is put on the dynamics of the population imbalance of the bosons, we compare the quantum evolution $\hat{S}_{z} (t)$ for the case $g_{IB} = 0$ to the classical dynamics $Z(t)$ in Fig.~\ref{ps} (b) and no discrepancies are observed among them. Hence, for the case $g_{IB} = 0$, we will always refer the classical $Z(t)$ dynamics as the quantum $S_{z}(t)$ evolutions. However, it should also be emphasized that the agreement between $\hat{S}_{z} (t)$ and $Z(t)$ takes place only for this non-interacting case. For $g_{IB} > 0$, on one side, the mixture has no classical mapping, on the other side, the quantum correlations among the bosons come into play, and, as a result, one can witness even a completely different quantum dynamics as compared to the classical one, albeit the fact that the bare Bose-Bose interaction always vanishes (see below).

The above classical interpretation provides us not only with a vivid picture for visualizing the quantum dynamics in a classical manner, but also with the profound physical insights with respect to its overall dynamical properties. In particular, the periodic motions for $\hat{S}_{y}(t)$ and $\hat{S}_{z}(t)$ obtained from the quantum simulations are a direct consequence of the integrability of the classical Hamiltonian $H_{cl}$. Owing to the energy conservation for the case $g_{IB} = 0$, $H_{cl}$ is completely integrable with all the corresponding classical trajectories, characterized by $[Z(t), \varphi(t)]$, being periodic in time \cite{PS}. Such an integrability is also transparently shown in the classical phase space [see Fig.~\ref{ps} (c)]. Depending on the initial condition, two distinguished types of motions are clearly observed: a periodic trajectory orbiting around the fix point either located at $(Z = 0,\varphi = 0)$ or $(Z = 0,\varphi = \pi)$, referred as the zero- and the $\pi$-phase mode for a 1D BJJ \cite{BJJ_2}.

\subsection{Breaking of the integrability} \label{Breaking_of_the_integrability}
In contrast to the above integrable limit, the presence of the impurity-Bose interaction leads to the energy transport between the two species and, hence, breaks the integrability for the bosonic species. In order to elaborate on this process in more detail, we decompose the interspecies interaction into various impurity-boson pair excitations 
\begin{equation}
\hat{H}_{IB} = \sum_{i,j = 0}^{\infty} \sum_{\alpha,\beta = L,R} U_{ij\alpha \beta} \hat{a}_{i}^{\dagger} \hat{a}_{j} \hat{b}_{\alpha}^{\dagger} \hat{b}_{\beta}, \label{H_IB_discrete}
\end{equation}
with $U_{ij\alpha \beta} = g_{IB} \int dx~ \phi_{i}(x) \phi_{j}(x) u_{\alpha}(x) u_{\beta}(x)$ and $\{\phi_{i}(x)\}$ being the single-particle basis for the DW potential. Moreover, $u_{L/R}(x)$, being the above mentioned localized Wannier-like states, are constructed via a linear superposition of the lowest two eigenstates $\phi_{0}(x)$ and $\phi_{1}(x)$. Note that Eq.~\eqref{H_IB_discrete} is obtained by means of an expansion of the field operator for the impurity $\hat{\psi}_{I}(x) = \sum_{i=0}^{\infty} \phi_{i}(x) \hat{a}_{i}$, meanwhile, by employing the two-mode approximation in Eq.~\eqref{2_mode_psi} for the bosonic species. Besides, all the eigenstate wavefunctions $\{\phi_{i}(x)\}$ are chosen to be real due to the preserved time-reversal symmetry in the single-particle Hamiltonian $\textit{h}_{\sigma}$.

Next, we group different pair excitations with respect to their bosonic indices as
\begin{align}
\hat{H}_{IB} &=\left[\sum_{i,j = 0}^{\infty} U_{ijLR} \hat{a}_{i}^{\dagger} \hat{a}_{j} \hat{b}_{L}^{\dagger} \hat{b}_{R} + \sum_{i,j = 0}^{\infty} U_{ijRL} \hat{a}_{i}^{\dagger} \hat{a}_{j} \hat{b}_{R}^{\dagger} \hat{b}_{L} \right] \nonumber \\
&+\left[ \sum_{i,j = 0}^{\infty} U_{ijLL} \hat{a}_{i}^{\dagger} \hat{a}_{j} \hat{b}_{L}^{\dagger} \hat{b}_{L} + \sum_{i,j = 0}^{\infty} U_{ijRR} \hat{a}_{i}^{\dagger} \hat{a}_{j} \hat{b}_{R}^{\dagger} \hat{b}_{R} \right]. \nonumber \\
\end{align}
By noticing the fact that
\begin{equation}
U_{ijLR} = U_{ijRL}, ~~U_{ijLL} = \eta U_{ijRR} \label{parity_sp}
\end{equation}
with $\eta = 1$ ($\eta = -1$) for $ n_{e,o} = |i - j|$ being an even (odd) number, together with the definitions given in Eq.~\eqref{spin_operators}, we finally arrive at 
\begin{align}
\hat{H}_{IB}&= \left(\sum_{i,j = 0}^{\infty} U_{ij}^{(1)} \hat{a}_{i}^{\dagger} \hat{a}_{j} \right) 2 \hat{J}_{x} +  \left(\sum_{|i-j| = n_e}^{\infty} U_{ij}^{(2)} \hat{a}_{i}^{\dagger} \hat{a}_{j} \right) \hat{N}_{B}  \nonumber \\
&+ \left(\sum_{|i-j| = n_o}^{\infty} U_{ij}^{(3)} \hat{a}_{i}^{\dagger} \hat{a}_{j} \right) 2 \hat{J}_{z}  \nonumber \\
&= \hat{H}_{IB}^{(1)} + \hat{H}_{IB}^{(2)}  + \hat{H}_{IB}^{(3)}. \label{H_IB_parity}
\end{align}
Here $U_{ij}^{(1)} =  U_{ijLR} = U_{ijRL}$, $U_{ij}^{(2)} =  U_{ijLL} = U_{ijRR}$ and $U_{ij}^{(3)} =  U_{ijLL} = -U_{ijRR}$. Let us emphasize that the Eq.~\eqref{parity_sp} relies on the fact that the DW potential is spatially symmetric, as a result, all its single-particle eigenstates $\{\phi_{i}\}$ respect the spatial parity symmetry. 

Equation \eqref{H_IB_parity} transparently elaborates how the interspecies interaction $\hat{H}_{IB}$ breaks the integrability for the Hamiltonian $\hat{H}_{B}$. Since both $\hat{H}_{IB}^{(1)}$ and $\hat{H}_{IB}^{(2)}$ commute with $\hat{H}_{B}$ [c.f. Eq.~\eqref{spin_BH_model}], it is the non-commutativity between $\hat{H}_{IB}^{(3)}$ and $\hat{H}_{B}$ that results in the energy non-conservation for the bosonic species, and breaks its integrability for $g_{IB} = 0$. Further inspecting the $\hat{H}_{IB}^{(3)}$ term in more detail, we notice that it corresponds to all the different single-particle excitations that reverse the impurity's spatial parity [see Fig.~\ref{ps} (a) for a schematic illustration]. With this, we conclude that those parity non-conservation transitions of the impurity leads to the integrability breaking for the majority bosonic species.

\subsection{Initial condition} \label{Initial_condition}
We prepare our impurity-Bose mixture initially as $|\Psi (0)\rangle = |\phi_{n} \rangle \otimes |\theta, \varphi \rangle$, being a product state between the two species. Here $|\phi_{n} \rangle$ is the $n$-th single-particle eigenstate for the impurity and $|\theta, \varphi \rangle$ denotes the ACS given by \cite{ACS_1,ACS_2} 
\begin{align}
|\theta, \varphi \rangle &= \frac{1}{\sqrt{N_{B}!}}\left[\text{cos}(\frac{\theta}{2}) \hat{b}^{\dagger}_{L} + \text{sin}(\frac{\theta}{2}) e^{i \varphi}\hat{b}^{\dagger}_{R}\right]^{N_B} ~|vac \rangle \nonumber \\
&= \sum_{N^{B}_{L}=0} ^{N_B} \left(\begin{array}{c} N_B \\ N^{B}_{L} \end{array} \right)^{1/2} \text{cos}^{N^{B}_L}(\theta/2)~\text{sin}^{N^{B}_{R}}(\theta/2) ~ e^{i N^{B}_{R} \varphi} ~|N^{B}_{L},N^{B}_{R} \rangle, \label{ACS}
\end{align}
which is the linear superposition of all the number states $\{|N^{B}_{L}, N^{B}_{R} \rangle\}$ and fulfills the completeness relation 
\begin{equation}
(N_{B}+1) \int \frac{d \Omega}{4 \pi} |\theta, \varphi \rangle \langle \theta, \varphi | = 1 \label{norm_ACS}
\end{equation} 
with $d \Omega = \text{sin}\theta d \theta d \varphi$ being the volume element. Physically, the ACS $|\theta, \varphi \rangle$ corresponds to the classical state $(Z, \varphi)$ in such a way that $\text{cos}\theta = (N^{B}_{L} - N^{B}_{R})/N_{B} = Z$ controls the initial population difference for the bosons and $\varphi$, possessing the same meaning with its classical counterpart, determines the phase difference between the two wells \cite{BJJ_4}. For a given ACS $|\theta, \varphi \rangle$, the mean values for the angular-momentum operators introduced in Eq.~\eqref{spin_operators} are \cite{BJJ_Rabi_3}
\begin{equation}
\langle\hat{S}_{x} \rangle = \text{sin}\theta\text{cos}\varphi, ~~~ \langle \hat{S}_{y}\rangle = \text{sin}\theta\text{sin}\varphi , ~~~\langle\hat{S}_{z}\rangle = \text{cos}\theta,  \label{initial_J_ACS}
\end{equation}
which satisfies the normalization condition $\langle\hat{S}_{x} \rangle^{2} + \langle\hat{S}_{y} \rangle^{2} + \langle\hat{S}_{z} \rangle^{2} = 1$. Together with the Eqs.~\eqref{EOM_Jyz} and \eqref{normal_S}, we conclude that, for the case $g_{IB} = 0$, the motion of the $\hat{\vec{S}}(t)$ vector starting from an arbitrary ACS always lies on a Bloch sphere with unit radius. Even for the case $g_{IB}>0$, where the vector $\hat{\vec{S}}(t)$ can jump out of the Bloch sphere significantly, the use of the ACS still allows us to visualize the quantum trajectory in a classical manner (see below), which simplifies the analysis of the complex quantum dynamics to a large extent, meanwhile, provides insights for the classical-quantum correspondence. Finally, let us note that the ACS has been implemented in recent ultracold experiments in a controllable manner. Tuning a two-photon transition between two hyperfine states of ${}^{87}\textrm{Rb}$ atoms, allows for preparing an ACS with arbitrary $|\theta, \varphi \rangle$ \cite{ACS_3,ACS_4}.  

In this paper, we aim at exploring the dynamical response of the majority bosonic species to the presence of the impurity. To this end, we quench at $t = 0$ the impurity-Bose interaction strength from initial $g_{IB} = 0$ to some finite value $g_{IB} > 0$, and monitor the quantum evolution of the bosonic population imbalance starting from a balanced population. While the initial state for the mixture is $|\Psi (0)\rangle = |\phi_{n} \rangle \otimes |\theta, \varphi \rangle$, without other specifications, we always choose the bosonic part being $|\theta, \varphi \rangle = |\pi/2, \pi/4 \rangle$. The corresponding $S_{z}(t)$ dynamics for this initial ACS and for the case $g_{IB} = 0$ has been detailed discussed above and is presented in Fig.~\ref{ps} (b) (red solid line). Furthermore, we also consider the scenarios for various initial impurity states $|\phi_{n} \rangle$, so as to explore its impact on the bosonic dynamics.

\section{Bosonic ITC transition} \label{Results}
\subsection{Onset of quantum chaos} \label{Onset_of_quantum_chaos}
\begin{figure}
  \centering
  \includegraphics[width=0.5\textwidth]{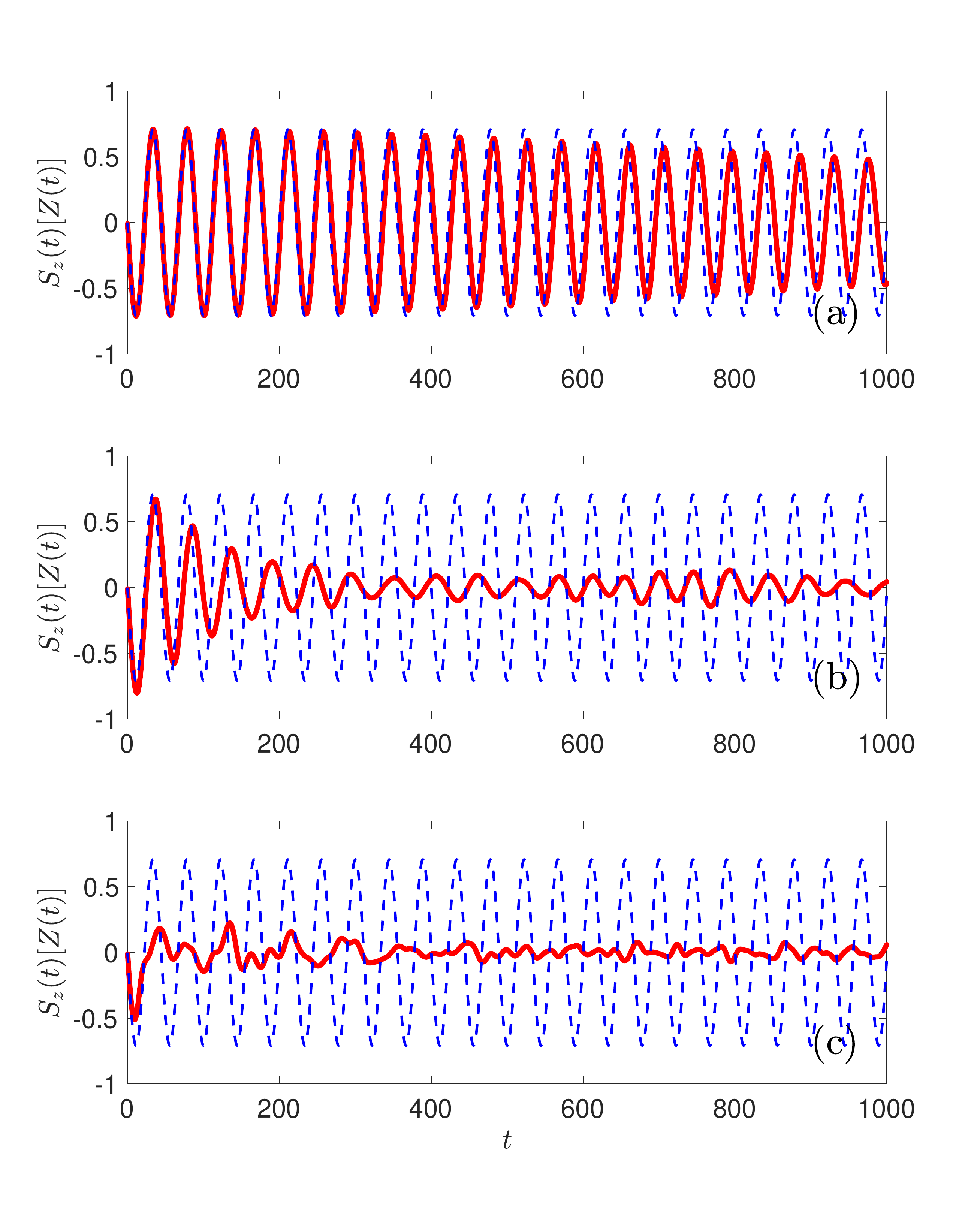}\hfill
  \caption{(Color online) Time evolution of the bosonic population imbalance $S_{z}(t)$ (red solid lines) for the initial state $|\Psi (0)\rangle = |\phi_{0} \rangle \otimes |\pi/2, \pi/4 \rangle$ and for various fixed impurity-Bose interaction strengths, in which (a) $g_{IB} = 0.01$, (b) $g_{IB} = 0.1$ and (c) $g_{IB} = 1.0$. For comparisons, the classical $Z(t)$ dynamcis is depicted as well (all blue dashed lines).}
\label{Jz_t_psiA_0}
\end{figure}

Let us first focus on the case where the impurity is initially prepared in its ground state. The many-body initial state for the mixture is then given by $|\Psi (0)\rangle = |\phi_{0} \rangle \otimes |\pi/2, \pi/4 \rangle$. Fig.~\ref{Jz_t_psiA_0} depicts the real-time population imbalance for the bosonic species $S_{z}(t)$  for various fixed postquench impurity-Bose interaction strengths $g_{IB} = 0.01$ [Fig.~\ref{Jz_t_psiA_0} (a)], $g_{IB} = 0.1$ [Fig.~\ref{Jz_t_psiA_0} (b)] and $g_{IB} = 1.0$ [Fig.~\ref{Jz_t_psiA_0} (c)], together with the classical $Z(t)$ dynamics (all blue dashed lines) which, as aforementioned, equivalents to the $S_{z}(t)$  for $g_{IB} = 0$. For a weak impurity-Bose interaction ($g_{IB} = 0.01$), the $S_{z}(t)$ dynamics is only slightly perturbed by the presence of the impurity, as a result, it leads to the small deviations of the population imbalance between the quantum and the classical simulations [c.f. Fig.~\ref{Jz_t_psiA_0} (a), red solid line and blue dashed line]. For a larger time scale ($t>5000$), a ``collapse-and-revival" behavior for $S_{z}(t)$ is observed (result is not shown here), manifesting its near integrability in this weak interacting regime \cite{BJJ_Rabi_3, BJJ_Driven_1}. Further increasing the interaction strength, the quantum $S_{z}(t)$ evolution becomes much more complicated and large discrepancies between $S_{z}(t)$ and ${Z}(t)$ are observed with respect to both the oscillation amplitude and the frequencies. For the case $g_{IB} = 1.0$, the quantum $S_{z}(t)$ dynamics finally becomes completely irregular [c.f. Fig.~\ref{Jz_t_psiA_0} (c), red solid line], signifying the onset of quantum chaos for the bosonic species. 

In order to diagnose such an ITC transition, meanwhile, to quantify the degree of the above observed quantum chaos, we employ the linear entanglement entropy (EE)
\begin{equation}
S_L = 1- \text{tr} \hat{\rho}_{1B}^{2} \label{linear_EE}
\end{equation}
for the bosonic species, which represents the bipartite entropy between the single boson and the $N_{B} -1$ bosons after tracing out the impurity \cite{QKT_7, QKT_8}. Here $\hat{\rho}_{1B}$ stands for the reduced one-body density matrix for the bosonic species \cite{dma1_1, dma1_2, BJJ_4}. Before proceeding, let us point out the reason for not using the spectral statistics as an indicator for the quantum chaos. Similar to the situation for a single particle in a 1D harmonic trap, the single DOF of the Hamiltonian $\hat{H}_{B}$ for a fixed particle number violates the Berry-Tabor conjecture, which states that the energy level spacing distribution follows the universal Poisson form for an integrable system \cite{level_1, level_2, level_3}. As a result, the variation of the level distribution for our mixture upon the increase of $g_{IB}$ can behave largely different as compared to other systems \cite{level_3}, and hence, it is insufficient to capture the quantum chaos. Upon a spectral decomposition of the reduced density matrix $\hat{\rho}_{1B}$, $S_L$ in Eq.~\eqref{linear_EE} can be expressed, with respect to the natural populations $\{n_{1}, n_{2}\}$, as $S_L = 1 - \sum_{i=1}^{2} n^{2}_{i}$. In this way, the linear EE alternatively measures the degree of the decoherence for the bosonic species. Note that the two-mode expansion employed in the Eq.~\eqref{2_mode_psi} renders the single-particle Hamiltonian $\textit{h}(x)$ being restricted to a two-dimensional Hilbert space and thus gives rise to only two natural populations obtained from the spectral decomposition \cite{BJJ_4}. For the case where all the bosons reside in the same single-particle state, the bosonic species is of complete coherence, as a result, we have $S_L = 0$. By contrast, for the case of maximal decoherence we have $n_{1} = n_{2} = 1/2$, which gives rise to the upper bound for the linear EE as $S_L = 1/2$.

\begin{figure}
  \centering
  \includegraphics[width=0.5\textwidth]{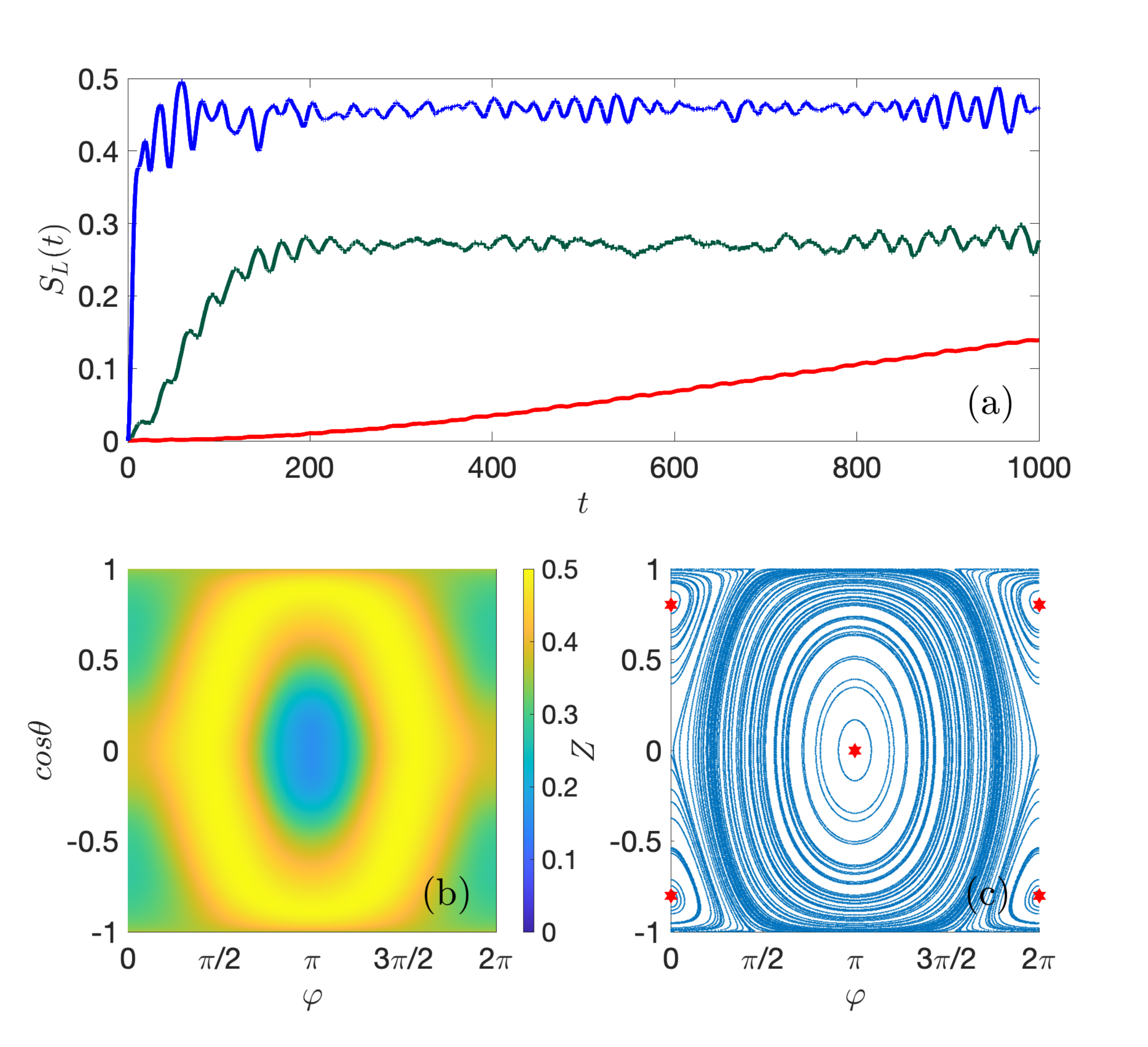}\hfill
  \caption{(Color online) (a) The linear EE evolutions for the initial state $|\Psi (0)\rangle = |\phi_{0} \rangle \otimes |\pi/2, \pi/4 \rangle$ and for the post-quench interaction strengths $g_{IB} = 0.01$ (red solid line), $g_{IB} = 0.1$ (green solid line) and $g_{IB} = 1.0$ (blue solid line). (b) Infinite-time averaged values for the linear EE for $g_{IB} = 1.0$ and for various ACSs. (c) A typical classical phase space for the BJJ with an attractive on-site interaction, where the red stars denote the corresponding classical fixed points.}
\label{EEs}
\end{figure}

The linear EE has been extensively used in the QKT systems as a signature of the quantum chaos \cite{QKT_7, QKT_8}. Depending on whether the corresponding classical trajectory is regular or chaotic, the linear EE behaves as either as rapidly growing or a slowly varying in a short time (referred to as the Ehrenfest time). On the other hand, the infinite-time averaged values of the linear EE for various initial ACSs additionally characterize the edge of the quantum chaos, denoted as the border between the integrable and the chaotic region in the corresponding classical phase space \cite{QKT_7, QKT_8}. Fig.~\ref{EEs} (a) reports the transient dynamics of the linear EE for the cases examined in Fig.~\ref{Jz_t_psiA_0} (a-c). At short times ($t<200$), the $S_L(t)$ evolution for a stronger interaction exhibits a more rapid growth as compared to the cases with a smaller $g_{IB}$. This is particularly obvious for the case $g_{IB}=1.0$, where we observe the linear EE surges to the value $S_L = 0.38$ at $t = 10$, while it only reaches to $S_L = 0.02$  ($S_L = 0.0007$) for the case $g_{IB}=0.1$ ($g_{IB}=0.01$). With this knowledge, we conclude that the different transient dynamical behaviors of the linear EE fully capture the ITC transition that is observed in the dynamics of the bosonic population imbalance. Besides, we shall also note that the linear EE for $t = 0$ trivially vanishes since all the bosons are initially condensed into the same single-particle state [c.f. Eq.~\eqref{ACS}].

Having investigated the transient dynamics of the linear EE for a specific ACS, let us now explore its asymptotic behaviors with respect to different ACSs, which, as aforementioned, characterize the edge of the quantum chaos. To this end, we compute the infinite-time averaged value of the linear EE (ITEE) for the initial state $|\Psi (0)\rangle = |\phi_{0} \rangle \otimes |\theta, \varphi\rangle$,
\begin{equation}
\overline{S}_L(\theta, \varphi) =lim_{T \rightarrow \infty}~ \frac{1}{T} \int_{0}^{T} dt~ S_{L}(t). \label{SL_bar}
\end{equation} 
Note that, the impurity initially always occupies the ground state $ |\phi_{0} \rangle$ and in our practical numerical simulations the time average is performed up to $t = 10^{4}$, being much larger than any other time scales involved in the dynamics. Before proceeding, let us point out the geometrical interpretation of the ITEE value. To show it, we first of all rewrite the linear EE in Eq.~\eqref{linear_EE} for time $t$ as \cite{QKT_7, QKT_8}
\begin{equation}
S_L(t) = \frac{1}{2} \left[1-  \sum_{\gamma = x,y,z} \langle\hat{S}_{\gamma} \rangle^{2}(t) \right],
\end{equation}
where we have used the relation 
\begin{equation}
\hat{\rho}_{1B} = \frac{1}{2} \left[1 + \sum_{\gamma = x,y,z}  \langle\hat{S}_{\gamma} \rangle \hat{\sigma}_{\gamma} \right],
\end{equation}
with $\{\hat{\sigma}_{\gamma} \}$ being the Pauli matrices. Since $S_L(t)$ is proportional to the instant distance of the vector $\hat{\vec{S}}(t) $ to the Bloch sphere, $\overline{S}_L$ thus measures its averaged distance for the entire dynamics. From the results in the QKT systems \cite{QKT_7, QKT_8}, we note that there exists a clear correspondence between the ITEE values and the classical phase space structure. Regions of low ITEE correspond to regular trajectories, while regions of high EE correspond to the chaotic trajectories. Moreover, a sudden change of the ITEE value takes place as one crosses the border between the integrable and the chaotic region, which, as aforementioned, characterizes the edge of the quantum chaos. Fig.~\ref{EEs} (b) depicts the computed ITEE values for various ACSs for the case $g_{IB} = 1.0$. Note that, we have rescaled the $\theta$ axis to $\text{cos}\theta$ since $\text{cos}\theta = Z$ [see discussions in Sec.~\ref{Initial_condition}]. 
Varying the initial ACS, the ITEE value varies accordingly. In particular, regions close to ($\text{cos}\theta = 0, \varphi = \pi$) and ($\text{cos}\theta = \pm 0.8, \varphi = 0, 2\pi$) possess significant low ITEE values as compared to the other places. Such a $\overline{S}_L(\theta, \varphi)$ profile significantly deviates from the structure of the non-interacting classical phase space. Instead, it bears a striking resemblance to the phase space with an attractive Bose-Bose interaction with the positions for those fixed points precisely match with those low ITEE regions [c.f. Fig.~\ref{EEs} (c), red stars]. Hence, we note that it indicates an effective Bose-Bose attraction existing among the bosons. In Sec.~\ref{Induced_Bose_Bose_attraction}, we will discuss this induced interaction in detail as well as its impact on the bosonic dynamics. 

\subsection{Recovery of the integrability} \label{integrable_dynamics}
In this section, we investigate the scenario where the impurity is initially pumped into a highly excited state. The out-of-equilibrium dynamics again is triggered by a sudden quench of the impurity-Bose interaction strength. Here, our main aim is to show that the integrability of the bosonic species is recovered by means of preparing the impurity in a highly excited state. The initial condition of the impurity, therefore, provides an additional DOF for controlling the ITC transition of the majority bosonic species. Here, we note that the employed notion of  ``integrability" specifically refers to how close the bosonic dynamics in the interacting cases ($g_{IB} > 0$) is to the one in the non-interacting integrable case ($g_{IB} = 0$), which is different from the commonly used context in which it is uniquely associated to the system's Hamiltonian.

\begin{figure}
  \centering
  \includegraphics[width=0.5\textwidth]{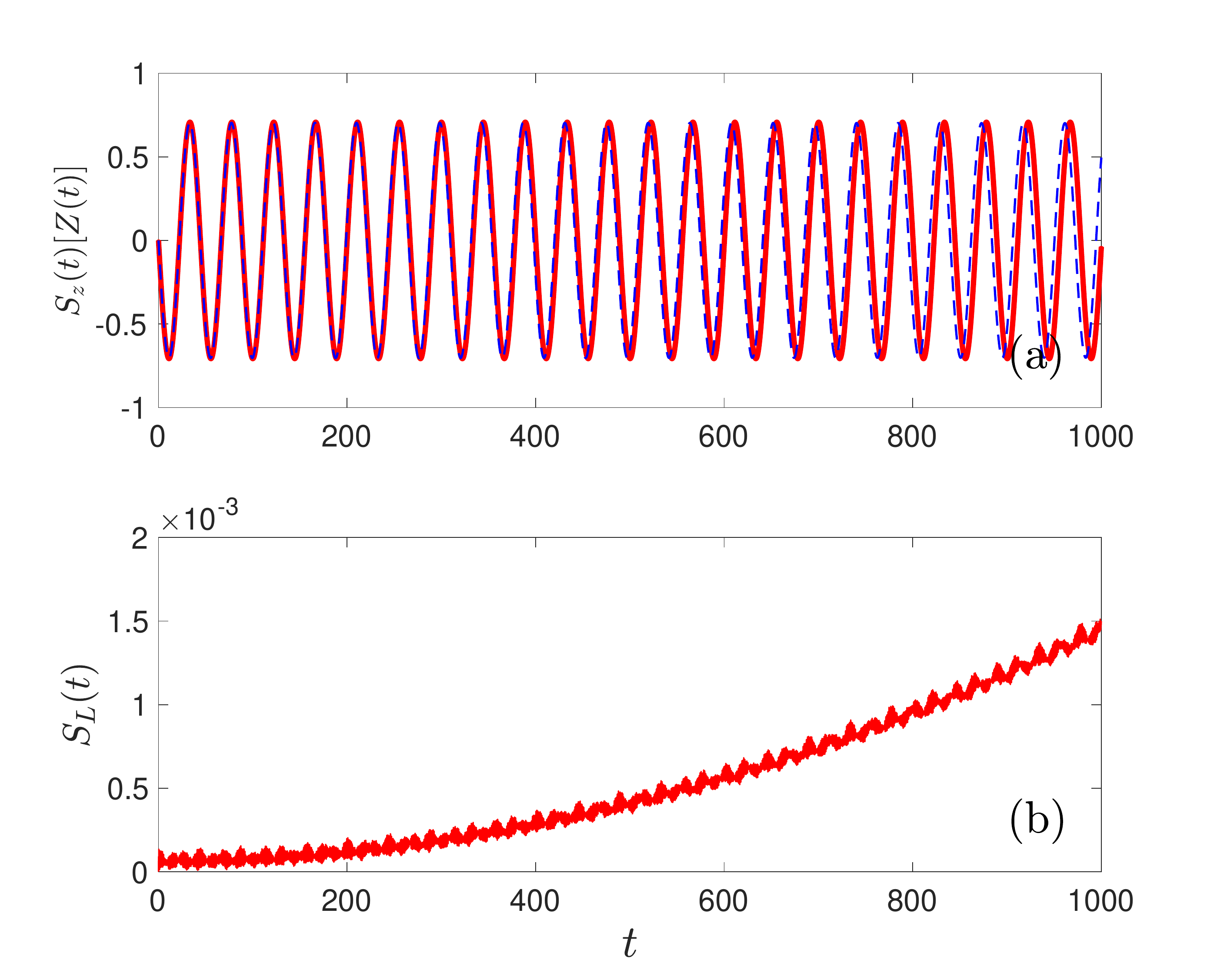}\hfill
  \caption{(Color online) Time evolution of the bosonic population imbalance $S_{z}(t)$ for $g_{IB} = 1.0$ and for the initial state $|\Psi (0)\rangle = |\phi_{150} \rangle \otimes |\pi/2, \pi/4 \rangle$ (red solid line), together with the classical $Z(t)$ dynamics (blue dashed line) which corresponds to the $S_{z}(t)$ dynamics for $g_{IB} = 0$. (b) The evolution of the linear EE for the corresponding case.}
\label{EEs_150}
\end{figure}

For an illustrative purpose, we consider the impurity is initially at $|\phi_{150} \rangle$, being the 150-th excited state, and focus on the case for the post-quench interaction strength $g_{IB} = 1.0$. The many-body state for $t = 0$ is again given by $|\Psi (0)\rangle = |\phi_{150} \rangle \otimes |\pi/2, \pi/4 \rangle$. The corresponding quantum evolution of the bosonic population imbalance $S_{z}(t)$ is depicted in Fig.~\ref{EEs_150} (a) (red solid line).  As compared to the classical $Z(t)$ dynamics [Fig.~\ref{EEs_150} (a), blue dashed line], we find a good agreement between them with negligible discrepancies. Interestingly, these discrepancies are even much smaller than the ones between $S_{z}(t)$ and $Z(t)$ for the case $g_{IB} = 0.01$ [c.f. Fig.~\ref{Jz_t_psiA_0} (a)]. Besides, we also note that the negligible increment of the corresponding linear EE in the course of the dynamics alternatively signifies the recovery of the integrability for the bosonic species [c.f. Fig.~\ref{EEs_150} (b)].

\section{Spectral analysis and induced interaction}\label{Discussions}
In order to shed light on the physics for the above-analyzed bosonic dynamics, hereafter, we perform a detailed spectral analysis for the mixture with respect to both the energy spectrum and the eigenstates via a numerically exact diagonalization (ED). In particular, we would like to unveil the physical origin for the observed ITC transition for the bosonic species manifested by the corresponding dynamics of the population imbalance. Moreover, we will discuss the presented Bose-Bose attraction induced by the impurity as well as its impact on the bosonic dynamics.

\subsection{Spectral structure}
Let us begin with the case for $g_{IB} = 0$. In the absence of the interspecies interaction, the two species are completely decoupled. As a result, the eigenenergy of the mixture is trivially given by $E = \epsilon_{k} + \epsilon^{B}_{l}$ with $ \epsilon_{k} $ and $ \epsilon^{B}_{l} $ being the $k$-th and $l$-th eigenvalue for the subsystem Hamiltonians $\hat{H}_{I}$ and $\hat{H}_{B}$, respectively. Owing to the neglected Bose-Bose interaction, the many-body spectrum for $\hat{H}_{B}$ is always equidistant with the energy difference $2J_{0}$ between the two successive levels, which accounts for the harmonic oscillation of the $S_z(t)$ dynamics for the case $g_{IB} = 0$ [c.f. Fig.~\ref{ps} (b)]. As for the impurity, due to the rapid growth of the energy difference between two successive eigenstates, the single-particle spectrum is inhomogeneous in which the high-energy part is much more sparse as compared to the low-energy one [c.f. Fig.~\ref{ps} (a)]. An important consequence for such a spectral structure on the mixture's many-body spectrum is the following. For $\delta_{i} > \Delta_B$, with $\delta_{i}  = \epsilon_{i+1} - \epsilon_{i} $ being the energy difference between the $i$-th and the $(i+1)$-th single-particle eigenstates for the DW potential (see also the discussions in Sec.~\ref{Hamiltonian}), and $\Delta_B$ representing the width of the spectrum for the Hamiltonian $\hat{H}_{B}$, a band-like structure is naturally formed in the high-energy part of the many-body spectrum with the band gap being $\delta_{i} -\Delta_B$, meanwhile, the energy levels within each band are equidistant.

\begin{figure}
  \centering
  \includegraphics[width=0.5\textwidth]{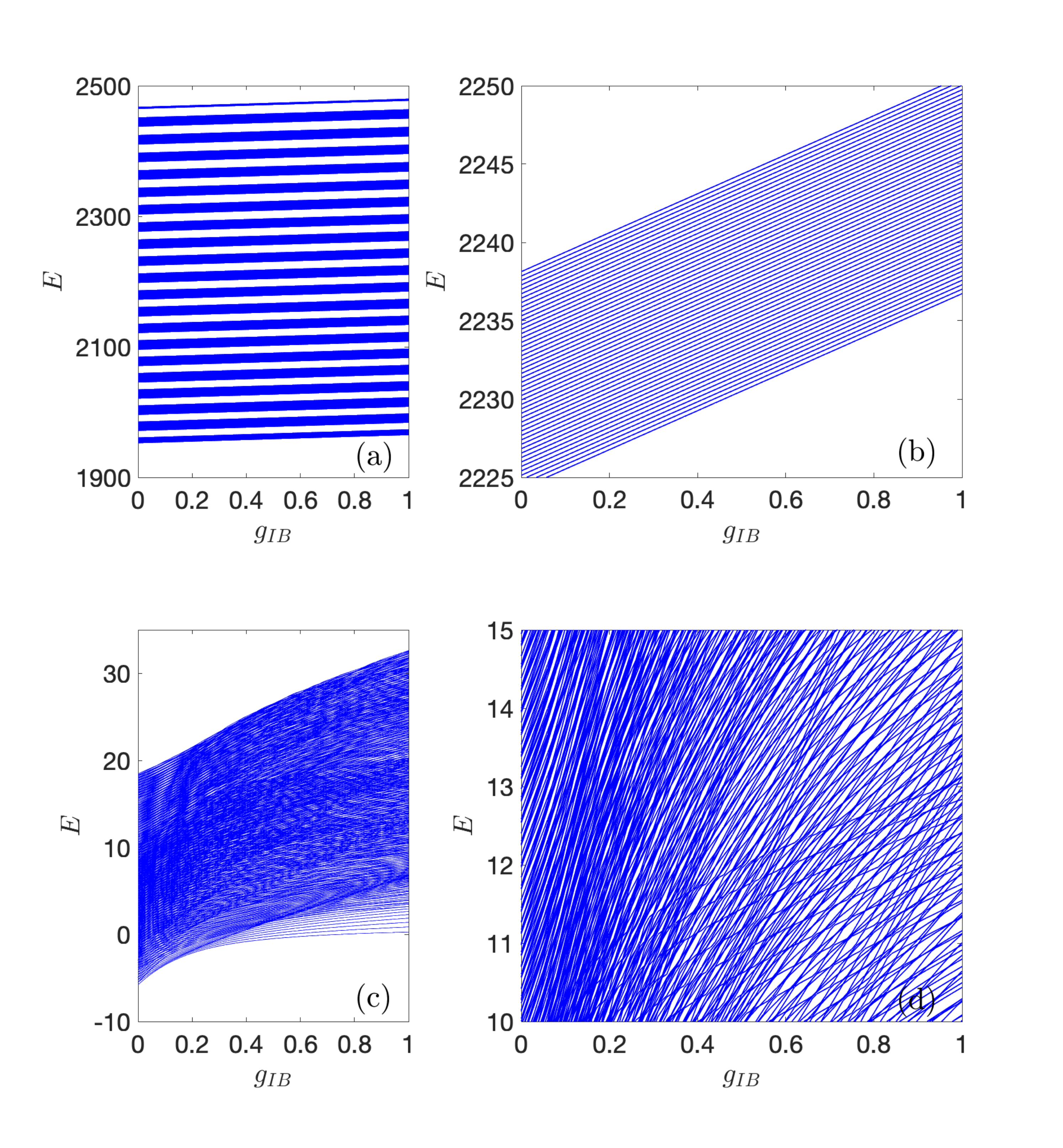}\hfill
  \caption{(Color online) Energy spectrum of the mixture as a function of impurity-Bose interaction strength $g_{IB}$. (a) High-energy part of the spectrum, (b) A zoom-in view of the high-energy spectrum, (c) Low-energy part of the spectrum, (d) A zoom-in view of the low-energy spectrum. }
  \label{spectrum}
\end{figure}

This simple picture, however, ceases to be valid upon the variation of the impurity-Bose interaction. Indeed, the inclusion of the interspecies interaction introduces additional coupling between the two subsystems and, as a result, our spectral analysis needs to be performed with respect to the complete mixture. Figure \ref{spectrum} showcases the many-body spectrum as a function of the interspecies interaction strength $g_{IB}$. Owing to the preserved spatial parity symmetry in the Hamiltonian $\hat{H}$, we present here only half of the spectrum which corresponds to the even parity eigenstates. With the increase of $g_{IB}$, the low-energy spectrum shows many avoided-crossings among the energy levels, which is in sharp contrast to the high-energy spectrum where only a linear growth of their values is observed [c.f. Figs.~\ref{spectrum} (a) and (c)]. Moreover, for the high-energy spectrum, features like the band-like structure as well as the equidistant energy levels within each band that are present in the non-interacting limit are retained in the interacting cases as well.

The above two distinguished spectral behaviors can roughly be understood via the structure of the impurity's single-particle spectrum  [c.f. Fig.~\ref{ps} (a)]. Owing to the large energy separations among those highly excited states, the transitions for the impurity among those states are significantly prohibited. From a many-body perspective, the resulting high-energy effective Hamiltonian of the mixture reads $\hat{H}^{\prime} = \hat{H}_{I} + \hat{H}_{B} +  \hat{H}_{IB}^{\prime}$, with 
\begin{align}
\hat{H}_{I}  &= \sum_{i\gg 1} \epsilon_{i}  \hat{a}_{i}^{\dagger} \hat{a}_{i}, ~~~~~~~~~~ \hat{H}_{B} = - 2J_0 \hat{J}_{x}, \nonumber \\
\hat{H}_{IB}^{\prime} &\approx \sum_{i \gg 1}2U_{i}^{(1)}  \hat{J}_{x} +   U_{i}^{(2)} \hat{N}_{B}  \approx  \sum_{i \gg 1} U_{i}^{(2)} \hat{N}_{B}. \label{H_high}
\end{align}
Here $U_{i}^{(1)} =  U_{iiLR} = U_{iiRL}$, $U_{i}^{(2)} =  U_{iiLL} = U_{iiRR}$ and we notice that $U_{i}^{(1)}  = g_{IB} \int dx~ \phi_{i}(x) \phi_{i}(x) u_{L}(x) u_{R}(x) \approx 0$, due to the negligible spatial overlap between the two localized states $u_{L}(x)$ and $u_{R}(x)$. Before proceeding, we note the validity condition for the above high-energy effective Hamiltonian as: $\delta_{i} \gg \epsilon_{IB}$ and $\delta_{i} \gg \Delta_{B}$ with $\epsilon_{IB}$ being the interspecies interaction energy per particle.  Eq.~\eqref{H_high} explains the observed high-energy spectral behaviors as follows: since the interspecies interaction $\hat{H}_{IB}^{\prime}$ now becomes the ``zero-point" energy of the mixture, the increment of the $g_{IB}$ thus only raises the energy level for those highly excited states. As a result, the band-like structure as well as equidistant nature that are formed in the non-interacting case are naturally preserved. 

In contrast, the densely distributed low-energy (single-particle) spectrum of the impurity facilitates the transitions among different (low-lying) many-body eigenstates caused by the interspecies interaction $\hat{H}_{IB}$ [c.f. Eq.~\eqref{H_IB_parity}]. With increasing $g_{IB}$, this results in the above observed avoided level-crossings among the low-energy many-body spectrum \cite{QKT_1}.
      
\subsection{Eigenstate delocalization}
The avoided level-crossings in the low-energy spectrum impact the characteristics of the corresponding eigenstates as well. Specifically, it results in a significant delocalization for those low-lying eigenvectors with respect to an integrable basis (see below), which, in turn, accounts for the chaotic nature of the bosonic non-equilibrium dynamics. To demonstrate this, we introduce the Shannon entropy  
\begin{equation}
S^{S}_{j} = -\sum_{k} c_{j}^{k} \ln{c_{j}^{k}} 
\end{equation}
for a many-body eigenstate $|\Phi_{j} \rangle$ of the mixture as a measure of the delocalization \cite{IPR_1, IPR_2}. Here $c_{j}^{k}  = |\langle\psi_{k}|\Phi_{j} \rangle|^{2}$ with $\{|\psi_{k}\rangle\} $ being the eigenbasis for the Hamiltonian $\hat{H}_{B}$ that are used as the ``integrable basis''. The Shanon entropy thereby measures the number of this integrable basis vectors that contribute to each eigenstate. As a result, the lower the Shanon entropy value is the closer this eigenstate $|\Phi_{j} \rangle$ is to a non-interacting eigenvector. From the random matrix theory (RMT), for a chaotic system described by the gaussian orthogonal ensemble (GOE), the amplitudes $c_{j}^{k}$ are independent random variables and all eigenstates are completely delocalized \cite{QKT_1}. However, due to the spectral fluctuations the weights $\{c_{j}^{k}\}$ fluctuate around $1/D$, yielding the averaged value $S_{\text{GOE}} = \ln{(0.48D)}$ \cite{IPR_1, IPR_2}. Here, we refer to $D = N_{B} +1$ as the Hilbert space dimension for the bosonic species, which is different from the single-species cases  \cite{IPR_1, IPR_2}.

\begin{figure}
  \centering
  \includegraphics[width=0.5\textwidth]{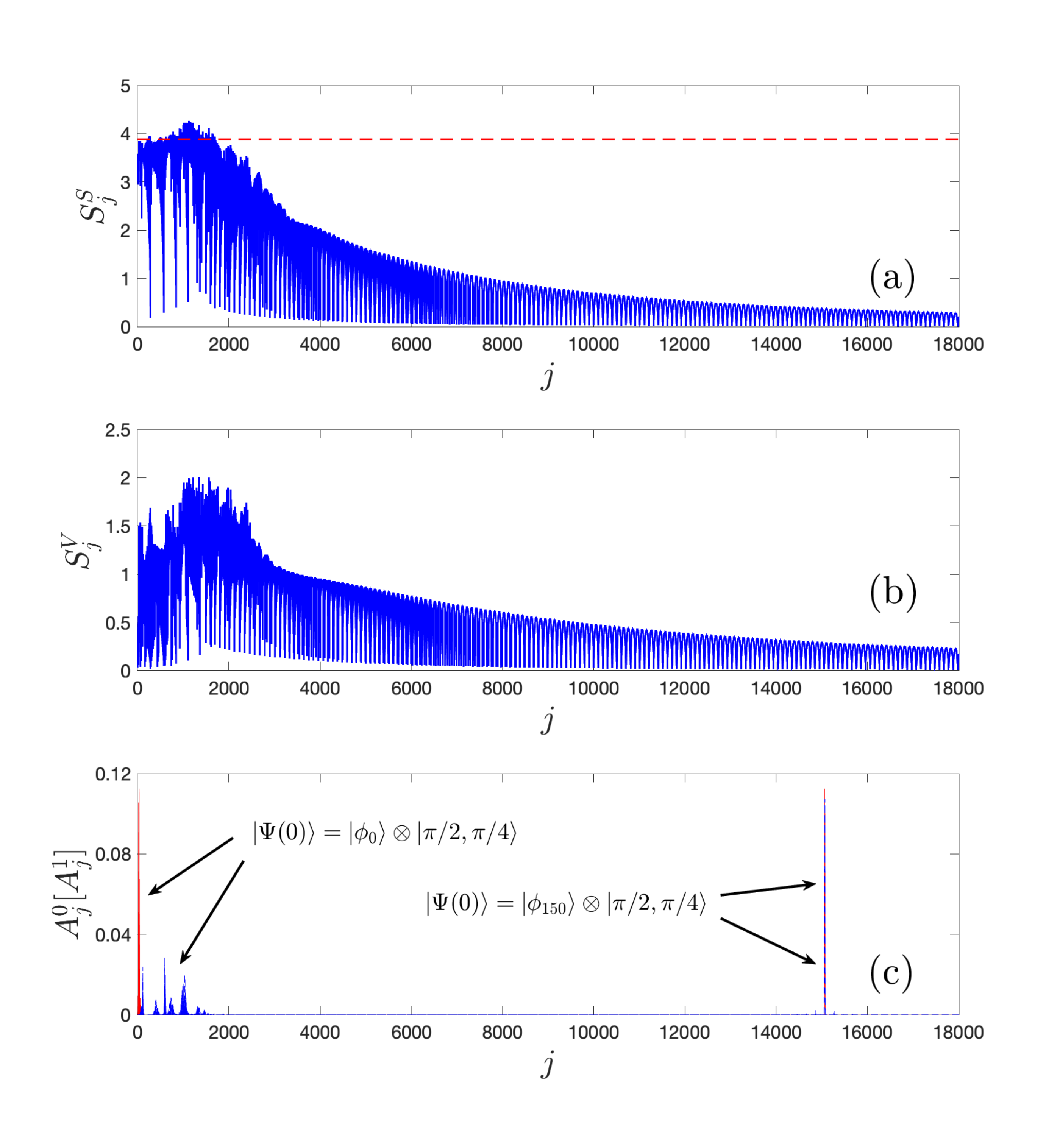}\hfill
  \caption{(Color online) (a) Shannon entropy $S^{S}_{j}$ for the many-body eigenstates as a function of quantum number $j$ for the case $g_{IB} = 1.0$. The red dashed line denotes the Shannon entropy from the GOE $S_{\text{GOE}} = 3.8812$. (b) Von Neumann entropy $S^{V}_{j} $ for the eigenstates for the case $g_{IB} = 1.0$. (c) Expansion coefficients $A_{j} = | \langle \Psi (0) | \Phi_{j} \rangle|^{2}$ with respect to eigenstates for initial states $|\Psi (0)\rangle = |\phi_{0} \rangle \otimes |\pi/2, \pi/4 \rangle$ (left part) and $|\Psi (0)\rangle = |\phi_{150} \rangle \otimes |\pi/2, \pi/4 \rangle$ (right part) and for the cases $g_{IB} = 0.0$ (red solid line and are denoted as $A_{j}^{0}$) and $g_{IB} = 1.0$ (blue dashed line and are denoted as $A_{j}^{1}$).}
  \label{IPR}
\end{figure}

Figure \ref{IPR} (a) presents the Shannon entropy of the many-body eigenstates as a function of their quantum numbers $j$ (sorted in the ascending order with respect to the energy) for the case $g_{IB} = 1.0$. The distinguished localization nature between the low-lying and the highly excited eigenvectors are clearly exhibited. While those low-energy eigenvectors are delocalized with the corresponding Shannon entropy values close to the result from the GOE $S_{\text{GOE}} = 3.8812$, for increasing $j$, a decrease of the $S^{S}_{j} $ value is clearly observed, indicating those high-energy eigenvectors are significantly localized. Thus, we may further conjecture that $S^{S}_{j} \rightarrow 0 $ for $j \rightarrow \infty$. Physically, the avoided level-crossings in the low-energy spectrum results in a strong mixing of different eigenstates with respect to their physical properties. In this way, an eigenstate from the non-interacting basis can be largely delocalized after experiencing a serious of avoided level-crossings \cite{QKT_1}. On the other hand, the localized nature for those high-lying excited states can also be readily seen from the effective Hamiltonian in Eq.~\eqref{H_high}. Since here $\hat{H}_{IB}^{\prime}$ corresponds to the ``zero-point " energy of the mixture, it is not surprising that the interacting basis (eigenstates of the mixture for $g_{IB}>0$) is similar to the non-interacting integrable basis.

Before proceeding, let us highlight that the degree of the localization for an eigenstate $|\Phi_{j} \rangle$ also reflects the degree of the encoded entanglement between the impurity and the majority bosons. To see this, we employ the von Neumann entropy for an eigenstate $|\Phi_{j} \rangle$ \cite{few_gs_7, Schmidt},
\begin{equation}
S^{V}_{j} = -\text{tr} (\hat{\rho}_{j}  \ln{\hat{\rho}_{j} } )
\end{equation}
with $ \hat{\rho}_{j}  = |\Phi_{j} \rangle \langle \Phi_{j} |$ being the corresponding density matrix. For the case where the two species are non-entangled, the eigenstate $|\Phi_{j} \rangle $ is simply of a product form with respect to the wavefunctions of the two species. Correspondingly, it gives rise to the von Neumann entropy $S^{V}_{j} = 0$. By contrast, any existing entanglement between the two species will lead to an increase of the von Neumann entropy, therefore, one can anticipate large $S^{V}_{j} $ values for those highly entangled eigenstates. The corresponding von Neumann entropies for various eigenstates for the case $g_{IB} = 1.0$ are shown in Fig.~\ref{IPR} (b). As compared to the Fig.~\ref{IPR} (a), a striking resemblance between the $S^{V}_{j}$ and $S^{S}_{j}$ distributions are transparently observed, manifesting the existence of the correspondence between a delocalized (localized) eigenstate to a large (small) von Neumann entropy value. Based on this knowledge, we refer to the above eigenstate delocalization as the entanglement induced delocalization.

Finally, let us discuss the impact of the eigenstate delocalization to the bosonic non-equilibrium dynamics. For the case $|\Psi (0)\rangle = |\phi_{0} \rangle \otimes |\pi/2, \pi/4 \rangle$, the initial state is mainly a linear superposition of those low-lying eigenvectors for both $g_{IB} = 0$ and $g_{IB} = 1.0$ [c.f. Fig.~\ref{IPR} (c), the left part]. Owing to the delocalization nature for the eigenstates of the mixture for large interspecies interactions, the expansion coefficients $\{A_{j}^{1}\}$ for $g_{IB} = 1.0$ are broadly distributed as compared to the ones ($\{A_{j}^{0}\}$) for $g_{IB} = 0.0$, reflecting the fact that much more eigenstates are involved in the bosonic dynamics. Since the energy levels for the interacting case are no longer equidistant, it thus gives rise to the completely irregular behaviors for the above $S_{z}(t)$ dynamics [c.f. Fig.~\ref{Jz_t_psiA_0} (c)]. In contrast, those highly excited states in the interacting basis preserve the main features of the non-interacting basis, leaving a similar distribution of the corresponding expansion coefficients [c.f. Fig.~\ref{IPR} (c), the right part]. Together with the equidistant nature for those high-lying energy levels, it thereby accounts for the integrable $S_{z}(t)$ motion for the initial state $|\Psi (0)\rangle = |\phi_{150} \rangle \otimes |\pi/2, \pi/4 \rangle$ and for the case $g_{IB} = 1.0$.

\subsection{Induced Bose-Bose attraction} \label{Induced_Bose_Bose_attraction}
The presence of the impurity not only brings the bosonic species into the chaotic regime, yielding an irregular behavior for the corresponding $S_{z}(t)$ motion, but also fundamentally changes its dynamical properties. As we will show below, the impurity effectively induces an attractive Bose-Bose interaction, which, in turn, leads to a completely different quantum trajectory as compared to the integrable case. To show it, we employ the time-averaged Husimi distribution (TAHD) \cite{TAHD_1, BJJ_4, QKT_7}
\begin{equation}
\overline{Q}_{H}(\theta, \varphi)  = lim_{T \rightarrow \infty} ~\frac{1}{ T } \int_{0}^{T} Q_{H} (\theta, \varphi,t) dt,
\end{equation}
with  
\begin{equation}
Q_{H} (\theta, \varphi,t) = \frac{N_B+1}{4\pi}  \langle \theta, \varphi | \hat{\rho}_{B}(t) | \theta, \varphi \rangle, \label{husimi_t}
\end{equation}
and $\hat{\rho}_{B}(t) $ being the reduced density matrix for the bosonic species after tracing out the impurity. According to the Eq.~\eqref{norm_ACS}, $Q_{H} (\theta, \varphi,t) $ satisfies the normalization condition $\int  Q_{H} (\theta, \varphi,t) d \Omega  = 1$. Physically, the TAHD represents the probability for the bosons to locate at a specific ACS $|\theta, \varphi\rangle$ averaged over the entire dynamics, which, with respect to its physical meaning, resembles to the probability density function (PDF) for a classical trajectory. In this sense, we note that the TAHD represents a quantum trajectory in an averaged manner.

\begin{figure*}
  \centering
  \includegraphics[width=1.0\textwidth]{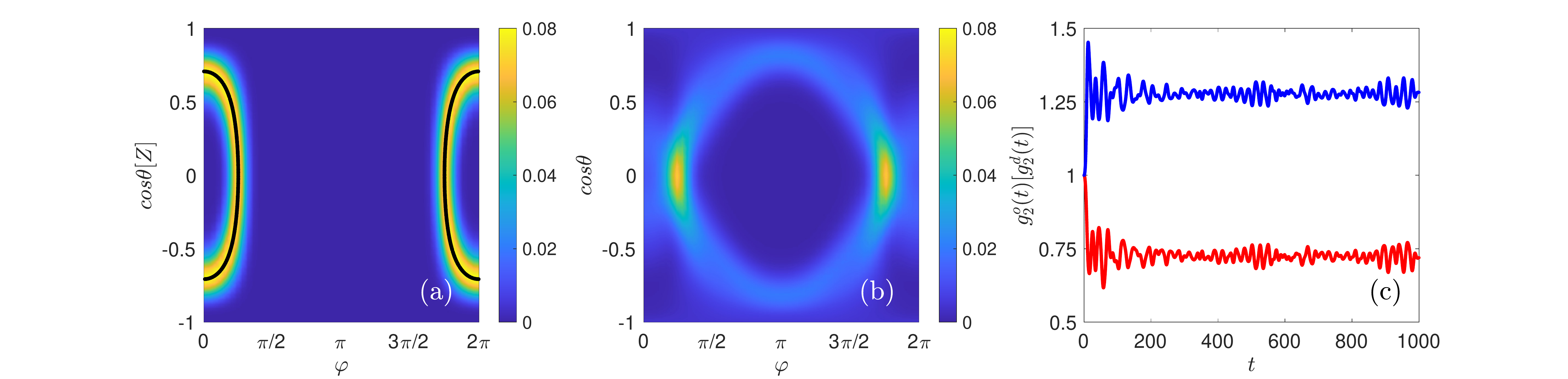}\hfill
     \caption{(Color online) Time-averaged Husimi distribution for the initial state $|\Psi (0)\rangle = |\phi_{0} \rangle \otimes |\pi/2, \pi/4 \rangle$ and for (a) $g_{IB} = 0.0$ and (b) $g_{IB} = 1.0$. Moreover, the black solid line in (a) denotes the classical trajectory starting from the phase point $(Z = 0, \varphi = \pi/4 )$. (c) The evolution of the pair-correlation function $g_2^{o}(t) $ (blue solid line) and $ g_2^{d}(t)$ (red solid line) for the case examined in (b). }
     \label{induce_int}
\end{figure*}

The computed TAHD for the initial state $|\Psi (0)\rangle = |\phi_{0} \rangle \otimes |\pi/2, \pi/4 \rangle$ and for the case $g_{IB} = 0$ is depicted in Fig.~\ref{induce_int} (a), together with the classical trajectory governed by the Hamiltonian $H_{cl} $ and starting from the phase point $(Z = 0, \varphi = \pi/4 )$ (black solid line). Compared to the classical trajectory, we note that the TAHD profile fully captures its main characteristic with those high $\overline{Q}_{H}(\theta, \varphi)$ regions precisely matching the positions for this classical trajectory, which additionally manifests the agreement between the quantum $S_{z}(t)$ and classical $Z(t)$ dynamics for the case $g_{IB} = 0$ [c.f. Fig. \ref{ps} (b)]. The TAHD for $g_{IB} = 1.0$, however,  deviates from the non-interacting case significantly and bears a striking resemblance to the classical trajectory corresponding to the BH Hamiltonian in Eq.~\eqref{BH_model} with an on-site attraction [c.f. Figs.~\ref{induce_int} (b) and \ref{EEs} (c)]. In this sense, we conjecture an effective Bose-Bose attraction is induced by the impurity in the dynamics which, in turn, alters the corresponding quantum trajectory.

This expectation is indeed confirmed by analyzing the pair-correlation function \cite{few_gs_6, few_gs_7,GPE_1}
\begin{equation}
g_2(\alpha, \beta) = \frac{\rho_{2}^{B}(\alpha, \beta)}{\rho_{1}^{B}(\alpha)\rho_{1}^{B}(\beta)},
\end{equation}
for the bosons, with $\rho_{2}^{B}(\alpha, \beta)$ and $\rho_{1}^{B}(\alpha)$ being the reduced two- and one-body density for the bosonic species and $\alpha, \beta = L,R$. Physically, $\rho_{2}^{B}(L, R)$ denotes a measure for the joint probability of finding one boson at the left well while the second is at the right well. Through the division by the one-body densities, the $g_2$ function excludes the impact of the inhomogeneous density distribution and thereby directly reveals the spatial two-particle correlations induced by the interaction \cite{few_gs_6, few_gs_7}. Based on this knowledge, let us first elaborate the $g_2$ function for the non-interacting case, which corresponds to the TAHD depicted in Fig.~\ref{induce_int} (a). Since there is no interaction among the particles, all the bosons thus can independently hop between the two wells, hence, it always results in $g_2^{o} = g_2^{d} = 1$, with $g_2^{o} = g_2(\alpha, \alpha)$ [$g_2^{d} = g_2(\alpha, \beta \neq \alpha)$] being the two-particle correlations within the same well (between the two wells). By contrast, the presence of the impurity-Bose interaction largely changes the above $g_2$ profile. As shown in Fig.\ref{induce_int} (c), the $g_2$ function quickly deviates from the initial values $g_2^{o} = g_2^{d} = 1$ to $g_2^{o} >1$ and $ g_2^{d} < 1$ for $t<5$ and persistently oscillate around their asymptotic values $g_2^{o} = 1.28$ and $g_2^{d} = 0.72$, respectively. Physically, such an evolution of the $g_2$ function indicates that the bosons are in favor of bunching together with a collective tunneling between the wells in the dynamics, which evidently manifests the existence of the Bose-Bose attraction induced by the impurity-Bose repulsion.

\section{Conclusions and Outlook} \label{Conclusions}

We have demonstrated that a non-interacting ultracold many-body bosonic ensemble confined in a 1D DW potential can exhibit a chaotic nature due to the presence of a single impurity. We trigger the non-equilibrium dynamics by means of a quench of the impurity-Bose interaction and monitor the evolution of the population imbalance for the bosons between the two wells. While the increase of the post-quench interaction strength always facilitates the chaotic motion for the bosonic population imbalance, it becomes regular again for the cases where the impurity is initially prepared in a highly excited state.  Employing the linear entanglement entropy, it not only enables us to characterize such an ITC transition but also implies the existence of an effective Bose-Bose attraction in the dynamics induced by the impurity. In order to elucidate the physical origin for the above observed ITC transition, we perform a detailed spectral analysis for the mixture with respect to both the energy spectrum as well as the eigenstates. In particular, two distinguished spectral behaviors upon a variation of the interspecies interaction strength are observed: while the avoided level-crossings take place in the low-energy spectrum, the energy levels in the high-energy spectrum possess the main features of the integrable limit. Consequently, it results in a significant delocalization for the low-lying eigenvectors which, in turn, accounts for the chaotic nature of the bosonic dynamics. In contrast, those highly excited states bear a high resemblance to the non-interacting integrable basis, rendering the recovery of the integrability for the bosonic species. Finally, we discuss the induced Bose-Bose attraction as well as its impact on the bosonic dynamics. 

Possible future investigations include the impact on the bosonic dynamics with the inclusion of several additional impurities and/or the bare Bose-Bose repulsion. Since for the latter there exists a competition between the bare Bose-Bose repulsion and the induced attractive interaction, this may significantly affect the bosonic ITC transition. Another interesting perspective is the study of the chaotic dynamics for an atomic mixture consisting of atomic species with different masses. The impact of the higher bands of the DW potential, beyond the two-site BH description for the bosonic species, is also an interesting perspective.

\begin{acknowledgments}
The authors acknowledge fruitful discussions with A. Mukhopadhyay and X.-B. Wei. This work has been funded by the Deutsche Forschungsgemeinschaft (DFG, German Research Foundation) - SFB 925 - project 170620586. K. K. gratefully acknowledges a scholarship of the Studienstiftung des deutschen Volkes. G. X. acknowledges support from the NSFC under Grants No. 11835011 and No. 11774316.
\end{acknowledgments}

\end{document}